\documentclass[%
aps,
pre,
superscriptaddress,
tightenlines,
showpacs,showkeys,
a4paper,
12pt,
longbibliography,
notitlepage
]{revtex4-1}
\usepackage[english]{babel}
\usepackage{amssymb,amsmath,stmaryrd,array}

\usepackage{graphicx}

\usepackage{subfig}
\makeatletter
\newcommand{\pdr}[2]{\frac{\partial #1}{\partial #2}}

\newcommand{\Tr}{\mathop{\rm Tr}\nolimits}

 \newcommand{\bs}[1]{\boldsymbol{#1}}
 \newcommand{\vc}[1]{\mathbf{#1}}
 \newcommand{\mvc}[1]{\mathbf{#1}}
 \newcommand{\uvc}[1]{\hat{\mathbf{#1}}}
 
 \newcommand{\ind}[1]{\mathrm{#1}}

\makeatother

\begin{document}
\DeclareGraphicsExtensions{.eps,.png,.pdf}
\title{%
Electrically assisted light-induced gliding of 
nematic liquid-crystal easy axis
at varying polarization azimuth 
of reorienting light: 
model and experiment 
}

\author{S.V.~Pasechnik}
\affiliation{%
Moscow State University of Instrument Engineering and Computer Science,
 Stromynka 20, 107846 Moscow, Russia}
\affiliation{%
Hong Kong University of Science and Technology, 
Clear Water Bay, Kowloon, Hong Kong}

\author{A.V.~Dubtsov}
\affiliation{%
Moscow State University of Instrument Engineering and Computer Science,
 Stromynka 20, 107846 Moscow, Russia}

\author{D.V.~Shmeliova}
\affiliation{%
Moscow State University of Instrument Engineering and Computer Science,
 Stromynka 20, 107846 Moscow, Russia}

\author{D.A.~Semerenko}
\affiliation{%
Moscow State University of Instrument Engineering and Computer Science,
 Stromynka 20, 107846 Moscow, Russia}

\author{V.G.~Chigrinov}
\email[Email address: ]{eechigr@ust.hk}
\affiliation{%
Hong Kong University of Science and Technology,
Clear Water Bay, Kowloon,
Hong Kong}

\author{M.A.~Sinenko}
\affiliation{%
Chernigov State Technological University,
 Shevchenko Street 95, 14027 Chernigov, Ukraine}

\author{Alexei~D.~Kiselev}
\email[Email address: ]{kiselev@iop.kiev.ua}
\affiliation{%
 Institute of Physics of National Academy of Sciences of Ukraine, 
prospekt Nauki 46, 03028 Kiev, Ukraine}
\affiliation{%
Hong Kong University of Science and Technology, 
Clear Water Bay, Kowloon, Hong Kong}

\date{\today}

\begin{abstract}
The phenomenological torque balance  model previously introduced to describe 
the electrically assisted light-induced gliding is generalized to 
study 
the reorientation dynamics of the nematic liquid crystal easy axis 
at photoaligned azo-dye films
under the combined action of in-plane electric field and 
reorienting UV light linearly polarized at varying polarization
azimuth, $\varphi_p$.
We systematically examine the general properties of 
the torque balance model by performing analysis of 
the bifurcations of equilibria at different
values of the polarization azimuth
and apply the model to interpret the experimental results.
In our experiments, it is found that,
in contrast to the case where $\varphi_p=0$ and
the light polarization vector is parallel to the initial
easy axis,
at $\varphi_p\ne 0$,
the pronounced purely photoinduced reorientation
is observed outside the interelectrode gaps.
It is also observed that,
in the regions between electrodes
with non-zero electric field,
the dynamics of reorientation
slows down with $\varphi_p$
and the sense of easy axis rotation
is independent of the sign of $\varphi_p$.
\end{abstract}

\pacs{%
61.30.Hn, 42.70.Gi
}
\keywords{%
nematic liquid crystal;  easy axis gliding; photo-alignment;
polarization azimuth
}
 \maketitle

%%%%%%%%%%%%%%%%%%%
\section{Introduction}
\label{sec:intro}
%%%%%%%%%%%%%%%%%%

It is well known that
external electric, magnetic or light
fields may produce deformations
of liquid crystal (LC) orientational structures 
initially stabilized by anisotropic 
boundary surfaces (substrates). 
There is a variety of 
related Fr\'{e}edericksz-type effects
that are at the heart of
operation of the vast majority of 
modern liquid crystal devices~\cite{Yang:bk:2006,Chigr:1999}.

Among the key factors 
that have a profound effect on behavior of 
the field induced orientational transitions
are the boundary conditions at the substrates. 
These conditions are determined by the anchoring characteristics
such as the anchoring energy strengths and the
\textit{easy axis},
$\mathbf{n}_e$, giving the direction of
preferential orientation of LC molecules at the surface.

In contrast to the traditional description of
Fr\'{e}edericksz-type transitions,
it turned out that 
reorientation processes induced by external fields
may additionally 
involve slow rotation of the easy axis.  
Over the past few decades this slow motion~---~the so-called 
\textit{easy axis gliding}~--~has received much 
attention as a widespread phenomenon
observed in a variety of liquid crystals
on amorphous glass~\cite{Oliveira:pra:1991},
polymer~\cite{Vetter:jjap:1993,Vorflusev:apl:1997,Faetti:epjb:1999,Lazarev:lc:2002,Janossy:pre:2004,
Joly:pre:2004,Faetti:pre:2005,Faetti:lc:2006,Pasechnik:lc:2006,Janossy:pre:2010}   
and solid~\cite{Faetti:epjb:1999,Pasechnik:lc:2006} 
substrates. 

Slow reorientation of the easy axis
also takes place on the photosensitive layers prepared using 
the photoalignment (PA) technique. These include
%such as 
poly-(vinyl)-alcohol (PVA) coatings
with embedded azo-dye molecules~\cite{Vorflusev:apl:1997},
polymer compound poly (vinyl methoxycinnamate)~\cite{Lazarev:lc:2002},
and the azo-dye films~\cite{Pasechnik:lc:2006}.

The PA  technique 
is employed in the manufacturing process of liquid crystal displays 
for fabricating high quality aligning substrates
and uses linearly polarized ultraviolet (LPUV) light
to induce anisotropy of the angular distribution
of molecules in azo-dye containing photosensitive
films~\cite{Chigrin:bk:2008}.
In this method, the easy axis is determined by 
the polarization azimuth of the pumping LPUV light,
whereas the azimuthal and polar anchoring
strengths may depend on a number of the governing parameters such as
the wavelength and the irradiation dose~\cite{Kis:pre2:2005}.

So, in a LC cell with the initially irradiated layer, 
subsequent illumination with reorienting light which polarization
differs from the one used to prepare the layer
can trigger the light-induced easy axis gliding. 
Such gliding  may be of considerable interest 
for applications such as LC rewritable devices~\cite{Chig:jjap:2008}
and 
the effects of the \textit{polarization azimuth}, $\varphi_p$, 
that characterizes orientation of the polarization vector of 
reorienting LPUV light, $\mathbf{E}_{UV}$
will be of our primary concern.

More specifically, we consider how the polarization azimuth 
affects the reorientational dynamics of the
electrically assisted light-induced azimuthal gliding of the easy
axis  that takes place
on photoaligned azo-dye layers
when irradiation of nematic LC (NLC)
cells with LPUV light is combined with the application 
of ac in-plane electric field~\cite{Pasechnik:lc:2008}.
It was observed that,
at certain combinations of the parameters
such as the amplitude of electric field  $E$, the light
intensity, $I_{UV}$, 
the exposure time, $t_{\mathrm{exp}}$, and the doze
of the initial UV irradiation, $D_p$, 
the switching off relaxation considerably slows
down up to few months.
The switching on dynamics of the gliding 
for both the linearly polarized and 
the nonpolarized reorienting light was studied 
in~\cite{Dubtsov:pre:2010}. 
In particular,
the results of the papers~\cite{Pasechnik:lc:2008,Dubtsov:pre:2010} 
demonstrate that the combined effect 
may be used as a tool to tune technical
parameters of LC memory devices. 

So, as compared to the case of 
purely light-induced reorientation of 
the easy axis governed by 
the effect of photoinduced ordering 
in azo-dye layers,
the dynamics of the electrically assisted
light-induced gliding can be additionally influenced by
the electric field, $E$.

In previous studies~\cite{Pasechnik:lc:2008,Dubtsov:pre:2010},
the reorienting light was linearly polarized along
the initial easy axis.
In this paper, 
our goal is to study 
how the polarization azimuth 
influences the surface mediated
reorientation processes that occur in NLC cells under 
the combined action of LPUV light and in-plane electric field. 

The layout of the paper is as follows.

In Sec.~\ref{sec:torque},
we formulate our phenomenological torque balance model of
electrically assisted photoinduced easy axis reorientation that takes
into account the effects due to the polarization azimuth.  Then, for
the dynamical system representing the model, the regime of
photosaturation is studied by analyzing the bifurcations of equilibria
at different values of the polarization azimuth.

In Sec.~\ref{sec:experiment}, 
after short description of the experimental procedure
used to measure the azimuthal easy axis angle as a function of
irradiation time, the theoretical curves computed from the model are
compared with the experimental data measured at various values of the
polarization azimuth.

Finally, in Sec.~\ref{sec:conclusion} we discuss the results
and make some concluding remarks.

%%%%%%%%%%%%
\section{Torque balance model}
\label{sec:torque}
%%%%%%%%%%%

In this section,
we carry out a theoretical investigation into 
the effects of the reorienting light polarization
by using a generalized version of
the phenomenological model
formulated in 
Refs.~\cite{Pasechnik:lc:2006,Pasechnik:lc:2008,Dubtsov:pre:2010}
to describe the effect of electrically assisted light-induced
azimuthal gliding.
According to this model,
the anchoring characteristics of the photoaligned layer
such as the easy axis, $\vc{n}_e$,
are determined by orientational order of  
LC molecules adsorbed by the azo-dye film. 

Initially, 
after PA treatment made before the cell is filled with LC,
ordering of the adsorbed molecules
is dictated by the orientational order parameter of the azo dye layer
induced by the initial irradiation with linearly polarized UV light,
$\vc{E}_{0}$. 
The initial direction of preferential orientation at the surface
$\vc{n}_{0}$
is typically normal to
the principal axis of the photoinduced anisotropy
defined by  the polarization vector $\vc{E}_{0}$.

When the cell filled with LC is subsequently irradiated
by the linearly polarized reorienting light $\vc{E}_{UV}$ 
with the polarization azimuth which differs from 
the one used at the preparation stage for PA treatment of
the azo-dye layer, absorbed LC molecules 
undergo light-induced transitions.
These transitions occur due to
reorientation of azo-dye molecules interacting with the adsorbed LC layer.
Thus this is the dynamics of photoinduced reordering of azo-dye molecules
that underlies the effect of the reorienting light on 
the anchoring properties of the photoaligned layer~\cite{Kiselev:pre:2012}.  

The easy axis can also be influenced by applying 
 an in-plane electric field $\vc{E}$ in the direction 
 perpendicular to $\vc{n}_{0}$.   
This can be explained in terms
of adsorption-desorption processes taking place in the near-surface
layer. 

Initially, in the absence of electric field,
the surface director $\vc{n}_s$ characterizing
average orientation of LC molecules  in the near-surface layer 
is directed along the easy axis  $\vc{n}_{0}$. 
So, the adsorption-desorption
processes do not influence
the undisturbed angular distributions of LC molecules
in the absorbed and near-surface layers
which are initially identical. 

 An electric field $\vc{E}$ produces a twist deformation on the
 distance $\xi$ defined as the \textit{electric coherence length}: 
$\xi=\dfrac{1}{E}\sqrt{K_{t}/(\varepsilon_0\Delta\varepsilon)}$,
 % \begin{equation}
 %   \label{eq:coh_length}
 %   \xi=\frac{1}{E}\sqrt{K_{t}/(\varepsilon_0\Delta\varepsilon)},
 % \end{equation}
where $\Delta\varepsilon$ is 
the electric permittivity anisotropy, $K_{t}$ is the effective Frank elastic
 constant for the twist deformation. 
 % In our case, the electric field is $E=U/d\approx 2$~V/\mum\, and,
% for the liquid crystal mixture E7 with the twist elastic constant
% $K_{22}\approx 6.5\times 10^{-12}$~N
% and the dielectric anisotropy $\Delta\varepsilon\approx 13.7$,
% the electric coherence length $\xi$ 
% can be estimated at about $0.12$~\mum.
At $\Delta\varepsilon>0$,
under this electric-field-induced deformation, the surface director,
$\vc{n}_s$, inclines towards the electric 
field $\vc{E}\perp \vc{n}_{0}$.
The absorption-desorption processes involving exchange of molecules 
between the differently aligned (absorbed and near-surface) layers
will result in reorientation of absorbed molecules.
Note that, owing to low probability of adsorption-desorption events,  
noticeable changes may require very long periods of time.

In the phenomenological model~\cite{Pasechnik:lc:2006,Pasechnik:lc:2008}, 
orientation of the easy axis $\vc{n}_e$ characterized by the azimuthal
angle $\varphi_e$ is defined by the balance of  the two torques: 
the torque transmitted from the bulk by the near-surface layer 
and the viscous torque proportional to \textit{the specific viscosity of gliding} $\gamma_e$. 
For the surface director $\vc{n}_s$ with the azimuthal angle 
$\varphi_s$,
the analogues balance involves the torque arising 
due to deviation of the surface director from the easy
axes (it is proportional to the surface anchoring energy strength
$W_s$),  the torque transmitted from the bulk and the viscous torque
proportional to the \textit{surface viscosity} $\gamma_s$. 

The resulting system of balance torque equations
for the easy axis and surface director azimuthal angles, $\varphi_e$
and $\varphi_s$,
reads
\begin{subequations}
  \label{eq:torq-orig}
\begin{align}
&
  \label{eq:phi_e-orig}
 \gamma_e 
\frac{\partial \varphi_e}{\partial t}
=
  K_E\, (\pi/2-\varphi_s) -
\frac{W_e}{2}\, \sin 2(\varphi_e-\varphi_m),
\\
&
 \label{eq:phi_s-orig}
\gamma_s 
\frac{\partial \varphi_s}{\partial t}
=
  K_E\, 
(\pi/2-\varphi_s) -
 \frac{W_s}{2}\, 
\sin 2(\varphi_e-\varphi_s),
\end{align}
\end{subequations}
where 
%$\gamma_e$ is the specific viscosity of gliding,
%$\gamma_s$ is the surface viscosity,
$K_E$ is the electric field induced torque coefficient inversely
proportional to the electric coherence length $\xi$, 
% $W_s$ is the surface ancchoring energy strength,
$W_e$ is the effective anchoring parameter which defines the strength
of coupling between the easy axis $\mathbf{n}_e$ and the initial state of surface orientation
described by the vector $\mathbf{n}_0$.

An important additional parameter is the phase shift $\varphi_m$ that enters the
second term on the right-hand side of Eq.~\eqref{eq:phi_e-orig} and
depends on the polarization azimuth of the reorienting light.
In the field-free regime with $E=0$ and $K_E=0$,
equation~\eqref{eq:phi_e-orig} assumes 
the simplified form
 \begin{align}
\label{eq:phi_e-zero}
\frac{\partial \varphi_e}{\partial t}
= -
\frac{W_e}{2 \gamma_e}\, \sin 2(\varphi_e-\varphi_m),
\end{align}
giving the formula for the easy axis angle  
 \begin{align}
  \label{eq:torq-phie-sol}
  \tan(\varphi_e(t)-\varphi_m)=\tan(\varphi_e(0)-\varphi_m) \exp[-t/t_e],
\end{align}
where $t_e=\gamma_e/W_e$ is
the characteristic time  of purely photoinduced easy axis
reorientation,
that represents the solution of Eq.~\eqref{eq:phi_e-zero}.
This formula
describes the phase shift 
as the azimuthal angle characterizing
the photosteady orientation of the easy axis: 
$\varphi_e^{(\ind{st})}=\varphi_m$
($\varphi_e^{(\ind{st})}=\varphi_m+\pi/2$) at $W_e>0$
($W_e<0$).
Interestingly, in our recent paper~\cite{Kiselev:pre:2012},
it was shown that, under certain assumptions,
equation~\eqref{eq:phi_e-zero} can be derived from
the diffusion model of photoinduced reordering
in azo-dye films. In this case, 
it turned out that the phase shift 
$\varphi_m$ equals the angle
between the polarization vector of the reorienting
light, $\vc{E}_{UV}$, and the axis 
directed along the normal to $\vc{n}_0$. 

Note that the two special cases where 
$\varphi_m=0$ and  $\varphi_m=\pi/2$  
has been previously treated in  
Refs.~\cite{Pasechnik:lc:2006,Pasechnik:lc:2008,Dubtsov:pre:2010}.
In subsequent sections, the effects due to variations of the phase shift
will be of our primary interest.

\begin{figure}%[!tbh]
\centering
\subfloat[$w_s=0.1$]{
   \resizebox{69mm}{!}{\includegraphics*{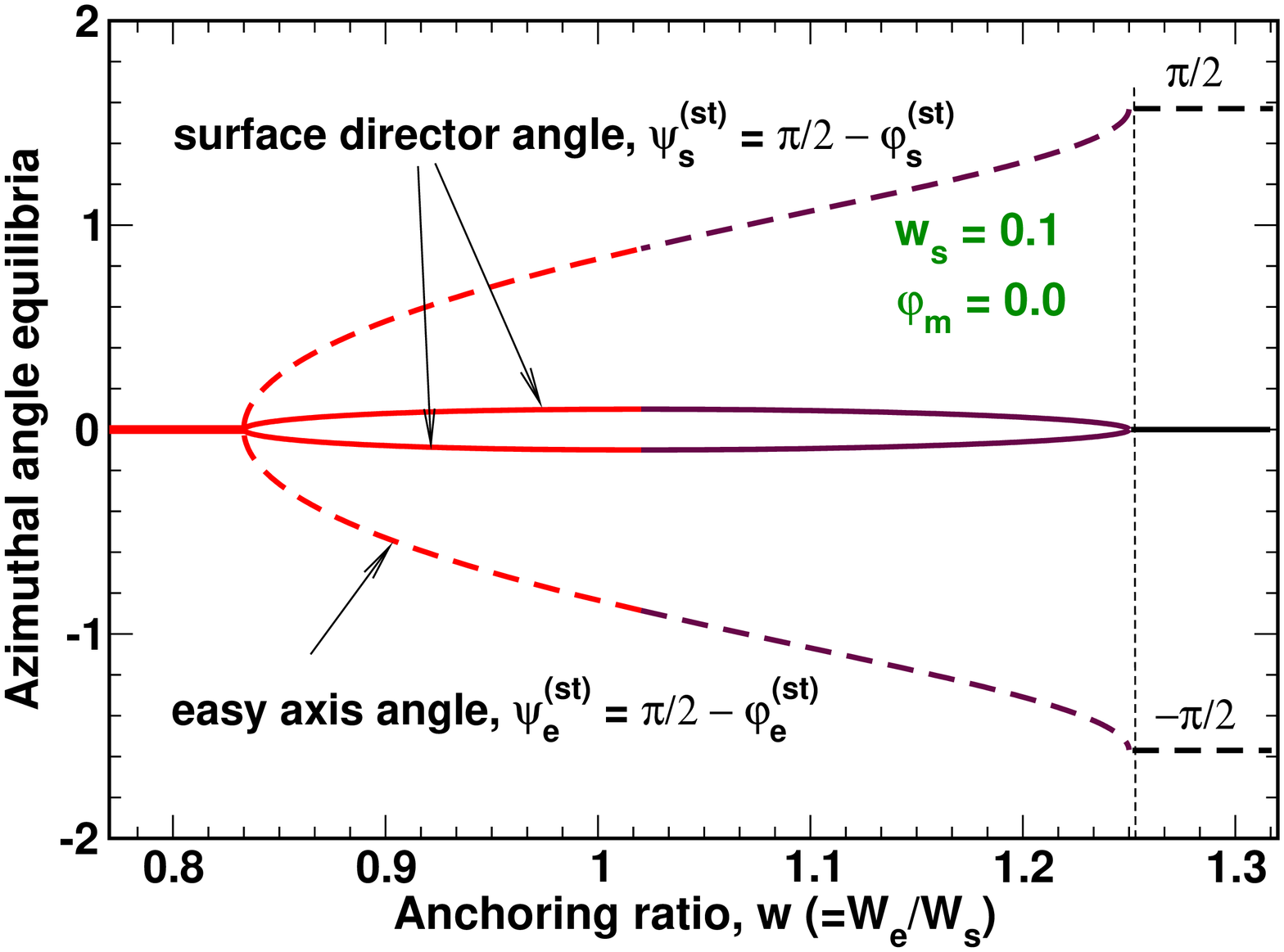}}
\label{subfig:ws-01}
}
\subfloat[$w_s=0.45$]{
   \resizebox{69mm}{!}{\includegraphics*{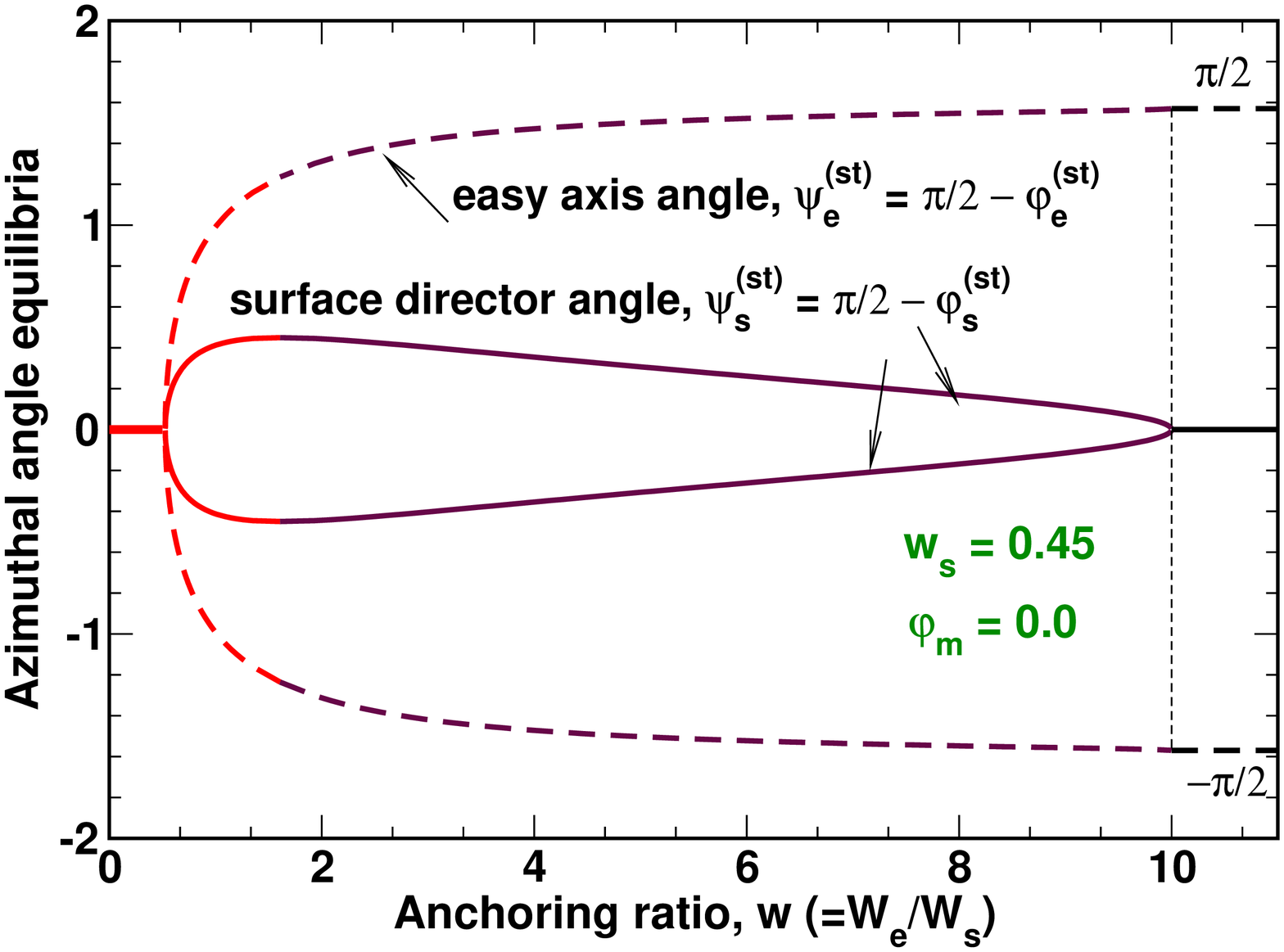}}
\label{subfig:ws-045}
}
\\
\subfloat[$w_s=0.49$]{
   \resizebox{69mm}{!}{\includegraphics*{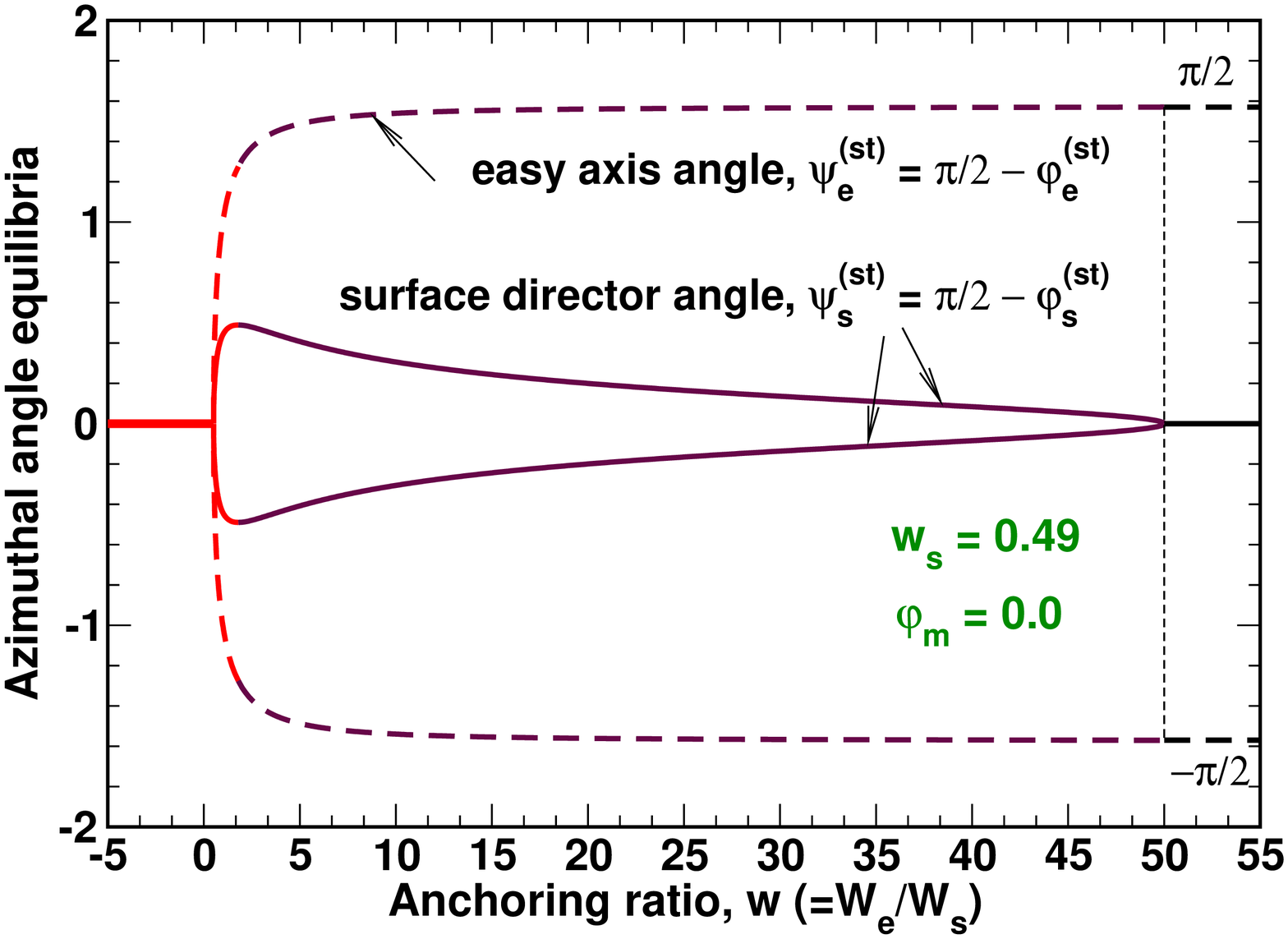}}
\label{subfig:ws-049}
}
\subfloat[$w_s=1.0$]{
   \resizebox{69mm}{!}{\includegraphics*{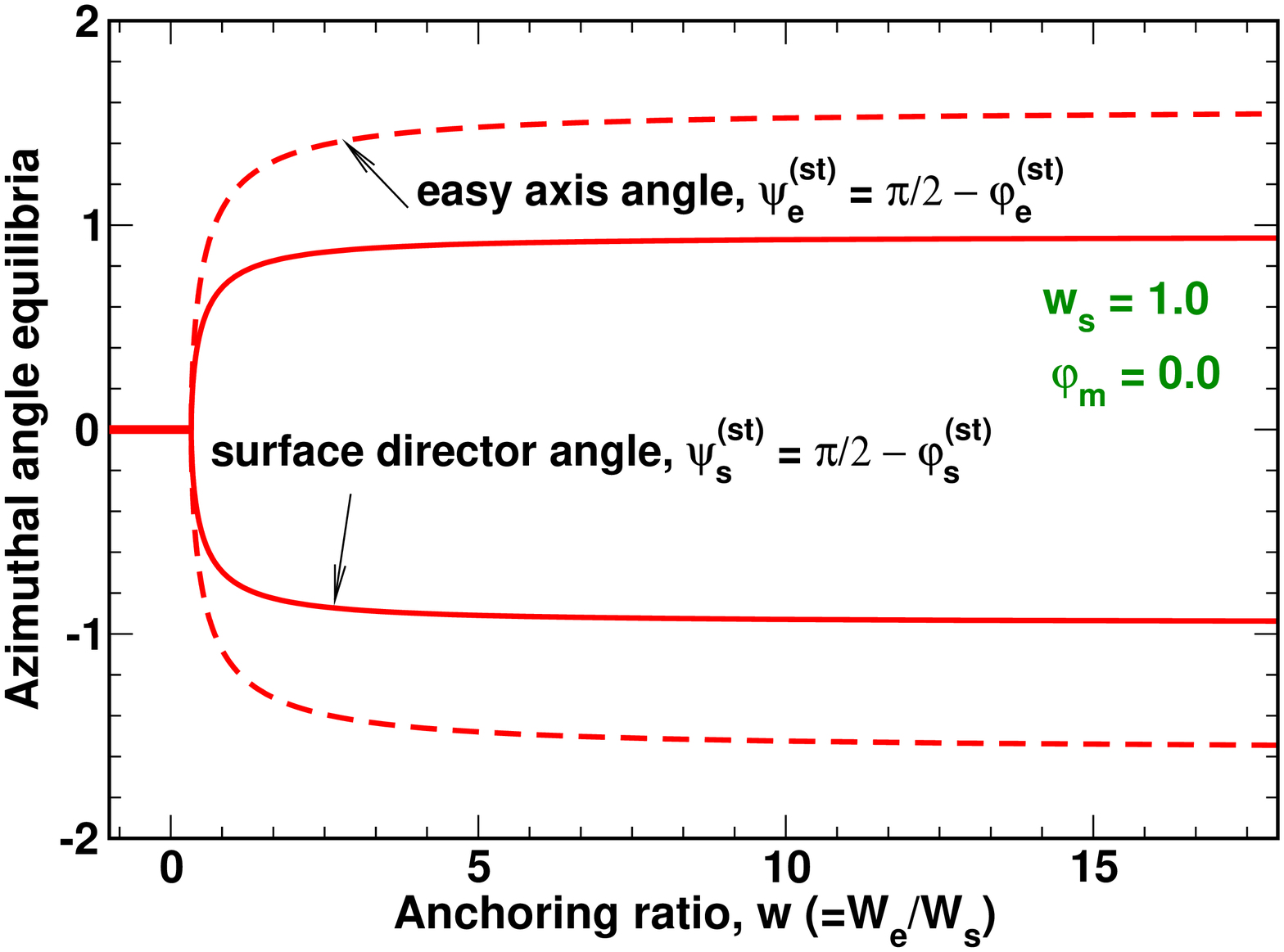}}
\label{subfig:ws-10}
}
\caption{%
Bifurcation diagram for equilibria
of easy axis and surface director azimuthal angles
as a function of the anchoring ratio, 
$w=W_e/W_s$, at  $\phi_m=0$.
Four cases are shown:
(a)~$w_s=0.1$;
(b)~$w_s=0.45$;
(c)~$w_s=0.49$;
and (d)~$w_s=1.0$.
}
\label{fig:bifur-ws}
\end{figure}

%%%%%%%%%%%%
\subsection{Bifurcation analysis of dynamical system}
\label{subsec:dyn-torque}
%%%%%%%%%%%

Before making comparison between the model and experiment, 
we shall dwell briefly on the general properties of 
the dynamical system~\eqref{eq:torq-orig}
which can be conveniently recast
into the following dimensionless form: 
\begin{subequations}
  \label{eq:torq-system}
\begin{align}
&
  \label{eq:phi_e}
\pdr{\varphi_e}{\tau}=
  \psi_s -
w_e \sin 2(\varphi_e-\varphi_m)\equiv \Phi(\varphi_e,\psi_s),
\quad
\tau=t/\tau_e,
\\
&
 \label{eq:psi_s}
\pdr{\psi_s}{\tau}=
\gamma
\Bigl\{  
-\psi_s +
w_s \sin 2(\varphi_e+\psi_s)
\Bigr\}
\equiv \gamma \Psi(\varphi_e,\psi_s),
\quad
\psi_s\equiv \pi/2-\varphi_s,
\end{align}
\end{subequations}
where $\tau_e=\gamma_e/K_{E}$ 
is the characteristic time  of 
easy axis reorientation;
$\gamma=\gamma_e/\gamma_s$ 
is the viscosity ratio;
$w_{e}=W_{e}/(2K_{E})$ 
and
$w_{s}=W_{s}/(2K_{E})$ 
are
the dimensionless anchoring parameters. 
The two symmetry relations
\begin{subequations}
  \label{eq:torq-symm}
\begin{align}
&
  \label{eq:torq-symm-a}
  \varphi_m\to\varphi_m+\pi/2\Longleftrightarrow w_e\to -w_e,
\\
&
  \label{eq:torq-symm-b}
\varphi_m\to-\varphi_m\Longleftrightarrow \{\varphi_e,\psi_s\}\to \{-\varphi_e,-\psi_s\}
\end{align}
\end{subequations}
describe how the system~\eqref{eq:torq-system} 
transforms when 
the phase shift is translated by $\pi/2$
(the polarization vector 
is rotated by $\pi/2$ around the $z$ axis which is normal to the layer)
and the phase shift changes its sign
(the polarization vector is reflected with respect to the $y$ axis).

The equilibrium (photosteady) states characterized by
the azimuthal angles,
$\varphi_e^{(\ind{st})}=\pi/2-\psi_e^{(\ind{st})}$ 
and $\psi_s^{(\ind{st})}=\pi/2-\varphi_s^{(\ind{st})}$,
can be found as 
the stable (attracting) stationary points of the dynamical
system~\eqref{eq:torq-system}.
In this section we shall generalize the results
of Ref.~\cite{Dubtsov:pre:2010} where 
the equilibria have been studied 
in relation to the \textit{anchoring ratio} 
$w=w_e/w_s=W_e/W_s$.
In particular, our analysis enables us 
to relax the constraints
requiring the anchoring parameter $w_s$
to be small ($|w_s|<0.5$) and the phase shift
to be zero, $\varphi_m=0$.

%%%%%%%%%%%%
\subsubsection{Branches of stationary states}
\label{subsubsec:steady-torque}
%%%%%%%%%%%

Equations for the stationary solutions  
of the system~\eqref{eq:torq-system} 
can be written in the following form:
\begin{align}
&
\label{eq:torq-st-branch}
\begin{cases}
    \sin 2(\varphi_e+\psi_s)=\psi_s/w_s,
\quad
\cos 2(\varphi_e+\psi_s) = -\dfrac{\mu}{w_s}\,\sqrt{w_s^2-\psi_s^2},
\\
    \sin 2(\varphi_e-\varphi_m)=\psi_s/w_e,
\quad
\cos 2(\varphi_e-\varphi_m) = \dfrac{\nu}{w_e}\,\sqrt{w_e^2-\psi_s^2},
\end{cases}
  \end{align}
where  $\mu, \nu\in \{+1,-1\}$ are the indices numbering four different branches of the
stationary states.

Similar to Ref.~\cite{Dubtsov:pre:2010},
our task is to examine behavior of the stationary
solutions when the \textit{anchoring ratio} 
$w=W_e/W_s$ varies whereas
the anchoring parameter $w_s$ and the phase shift
$\varphi_m$ are both kept constant.
For this purpose,
it is convenient to recast
the stationarity equations~\eqref{eq:torq-st-branch}
into  the parametrized form  
\begin{subequations}
\label{eq:param-st-branch}
  \begin{align}
&
\label{eq:W_nu}
w =W_\nu(t;w_e)=
t^{-1}
\Bigl\{
t\cos2(t+\varphi_m)+\nu
\sqrt{w_e^2-t^2}\cos2(t+\varphi_m)\Bigr\},
\\
&
\label{eq:W1_mu}
w^{-1} =W_\mu^{(1)}(t;w_s)=
t^{-1}
\Bigl\{
t\cos2(t+\varphi_m)+\mu
\sqrt{w_s^2-t^2}\cos2(t+\varphi_m)\Bigr\},
    \end{align}
\end{subequations}
where the surface director azimuthal angle 
$\psi_s$ plays the role of a parameter $-|w_s| \le t\equiv
\psi_s\le |w_s|$.
Then
the solutions of the equation
\begin{align}
  \label{eq:D-mu-nu}
 \frac{1}{W_\mu^{(1)}(t;w_s)}=
W_\nu\Bigl(
t;w_s/W_\mu^{(1)}(t;w_s)
\Bigr)\Rightarrow
t\in D_{\nu\mu}
\end{align}
form the branch of stationary points $D_{\nu\mu}$
and we can use Eq.~\eqref{eq:W1_mu}
to derive the relations 
\begin{align}
  \label{eq:bifur-curv-psi-s}
  \psi_s^{(\ind{st})}(w)=
      \begin{cases}
        w= \dfrac{1}{W_\mu^{(1)}(t;w_s)}\\
        \psi_s^{(\ind{st})}=t 
      \end{cases},
\quad t\in D_{\nu\mu}
\end{align}
representing the dependence of the stationary values
of the surface director angle
on the anchoring ratio in the form of
the parametrized curve in the $w-\psi_s$ plane. 
Similarly, in the $w-\psi_e$ plane, 
the curve
\begin{align}
  \label{eq:bifur-curv-psi-e}
  \psi_e^{(\ind{st})}(w)=
      \begin{cases}
        w= \dfrac{1}{W_\mu^{(1)}(t;w_s)}\\
        \psi_e^{(\ind{st})}=t+2^{-1} \bigl[
\mu\arcsin(t/w_s)+(1-\mu)\pi/2
\bigr]
      \end{cases},
\quad t\in D_{\nu\mu},
\end{align}
describes the corresponding branch of stationary easy axis angles.

We now pass on to the stability analysis of the stationary states.
Following the standard approach~\cite{Kuznetsov:bk:1998,Gucken:bk:1990},
the criteria of linearized stability 
are formulated in terms of
the linearization matrix of the system~\eqref{eq:torq-system},
in the neighborhood of the fixed point
$(\varphi_e^{{(\ind{st})}},\psi_s^{{(\ind{st})}})$
\begin{align}
\label{eq:stab-matr-H}
        \mvc{H}=\begin{pmatrix}1& 0\\0 & \gamma\end{pmatrix}\cdot
        \bs{\Lambda},
\quad
\bs{\Lambda}=
-\begin{pmatrix}
\dfrac{\partial \Phi}{\partial \varphi_e} & \dfrac{\partial \Phi}{\partial \psi_s}
\\
\dfrac{\partial \Psi}{\partial \varphi_e} & \dfrac{\partial
  \Psi}{\partial \psi_s}
\end{pmatrix}\Biggr|_{\varphi_e=\varphi_e^{{(\ind{st})}},\; \psi_s=\psi_s^{{(\ind{st})}}},  
\end{align}
so that, in the case of two-dimensional systems,
the stability conditions are given by
\begin{align}
&
\label{eq:stab-cond}
\Tr{\mvc{H}}>0,\quad
\det{\mvc{H}}>0.
      \end{align}
The expression for the matrix $\bs{\Lambda}$ 
evaluated for the branch of stationary points $D_{\nu\mu}$
\begin{align}
&
\label{eq:stab-matr-Lmb}
        \bs{\Lambda}\bigr|_{t\in D_{\nu\mu}}\equiv \bs{\Lambda}_{\nu\mu}(t)=
        \begin{pmatrix}
          2\nu\sqrt{w_e^2-t^2} & -1\\
          2\mu\sqrt{w_s^2-t^2} & 1+2\mu\sqrt{w_s^2-t^2}
        \end{pmatrix}
      \end{align}
can be derived with the help of equations~\eqref{eq:torq-st-branch}.

\begin{figure}%[!tbh]
\centering
\subfloat[$\varphi_m=\pi/18$]{
   \resizebox{69mm}{!}{\includegraphics*{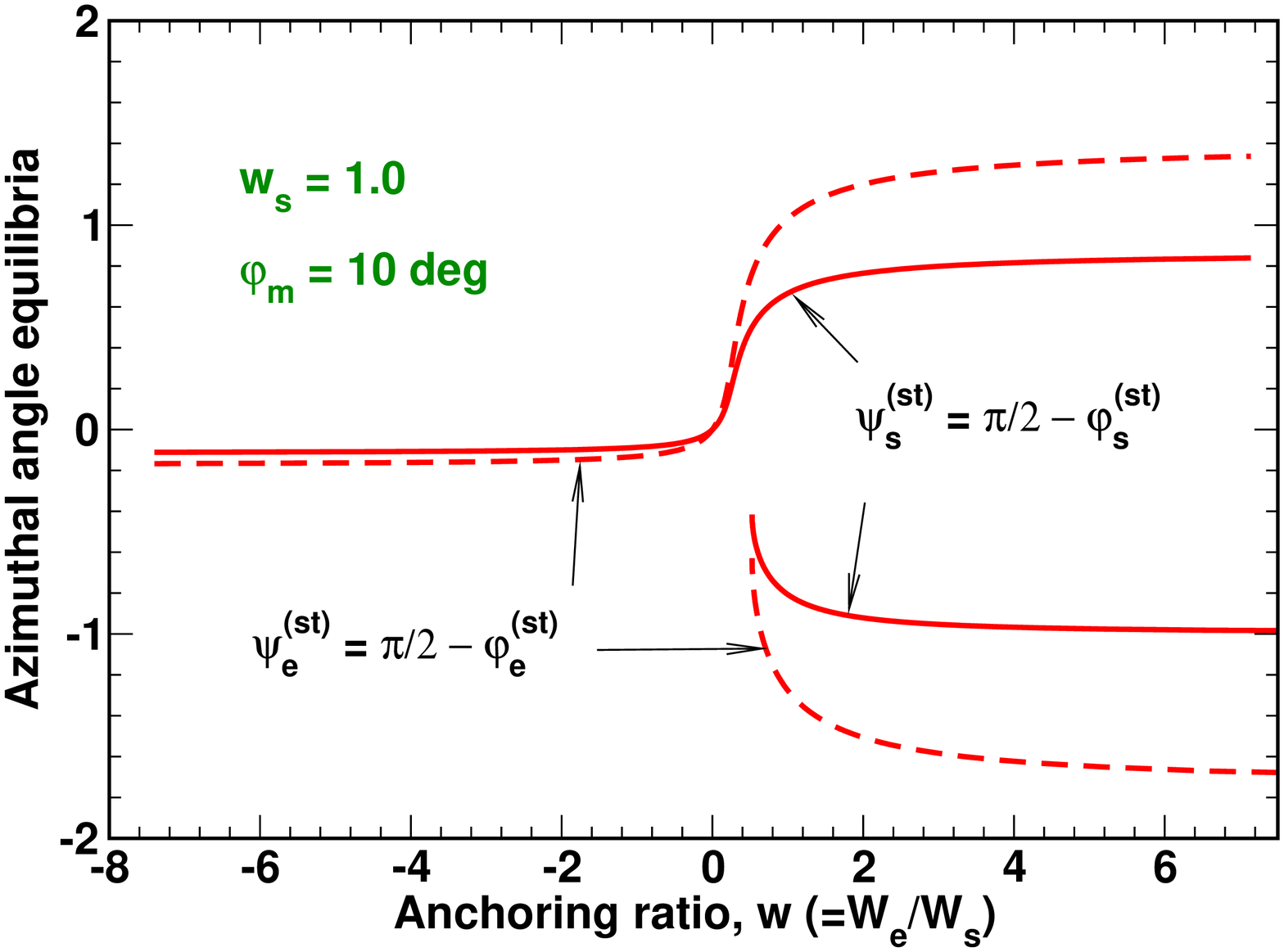}}
\label{subfig:phim-10}
}
\subfloat[$\varphi_m=\pi/9$]{
   \resizebox{69mm}{!}{\includegraphics*{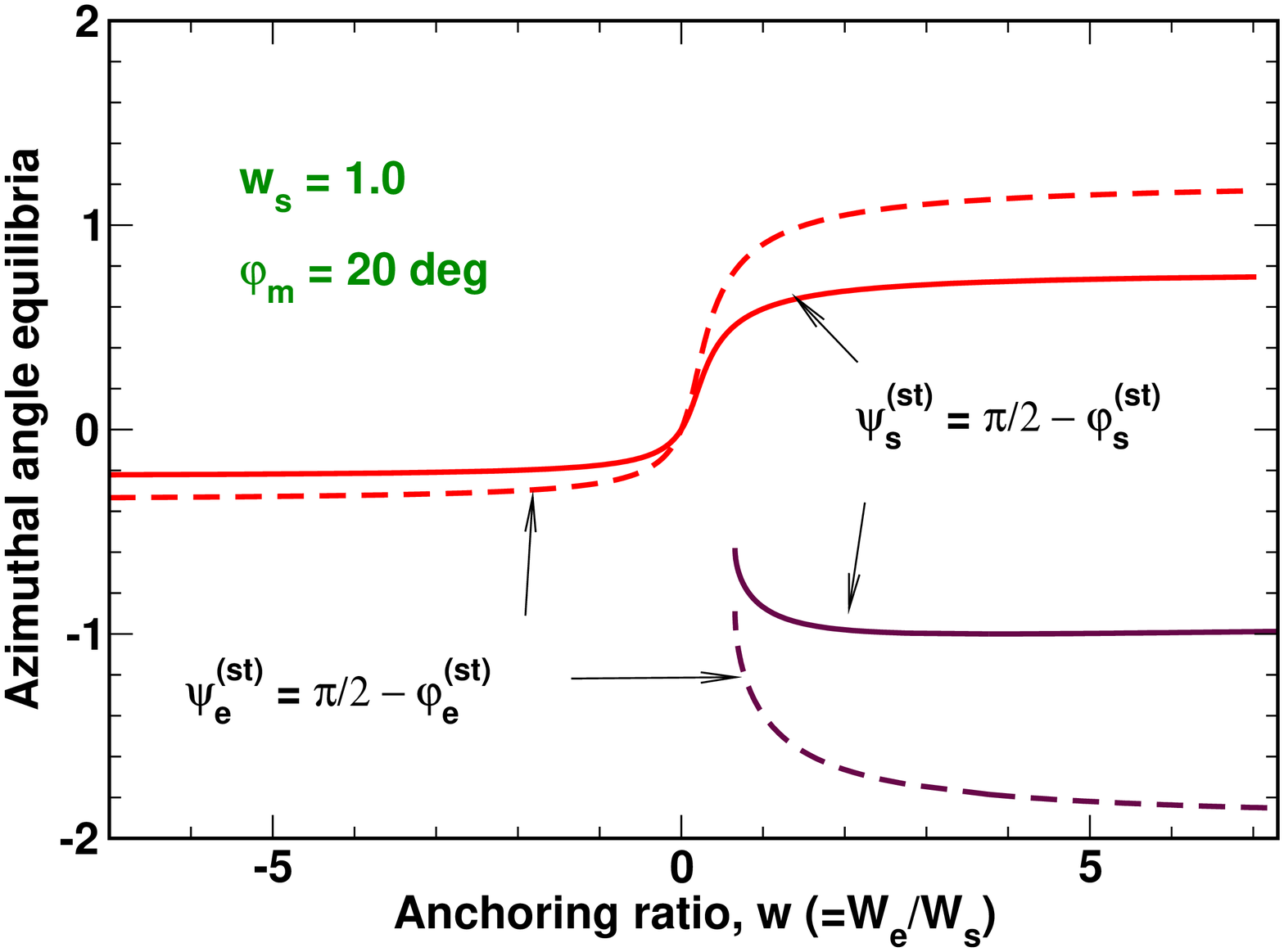}}
\label{subfig:phim-20}
}
\\
\subfloat[$\varphi_m=5\pi/36$]{
   \resizebox{69mm}{!}{\includegraphics*{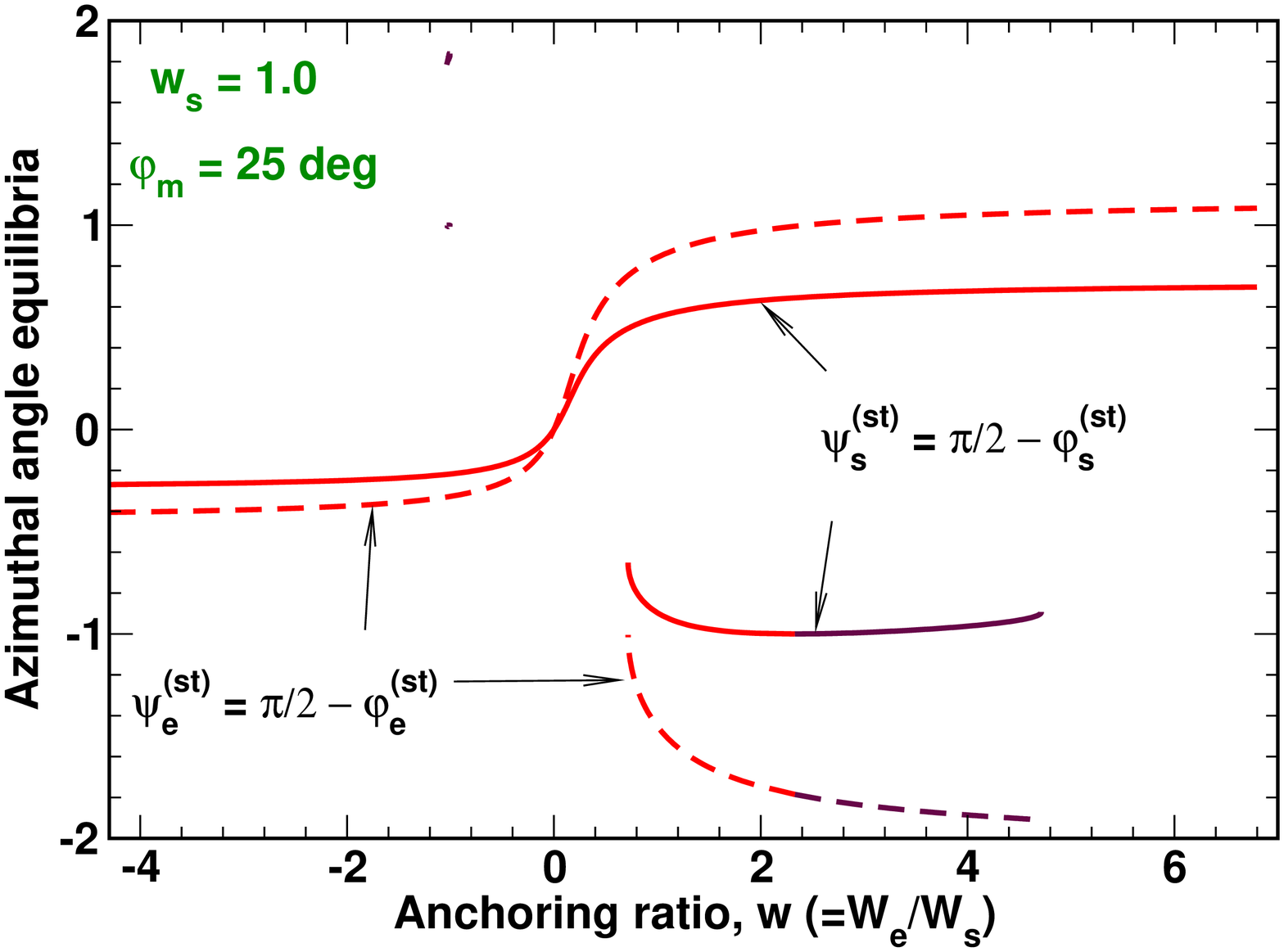}}
\label{subfig:phim-25}
}
\subfloat[$\varphi_m=\pi/4$]{
   \resizebox{69mm}{!}{\includegraphics*{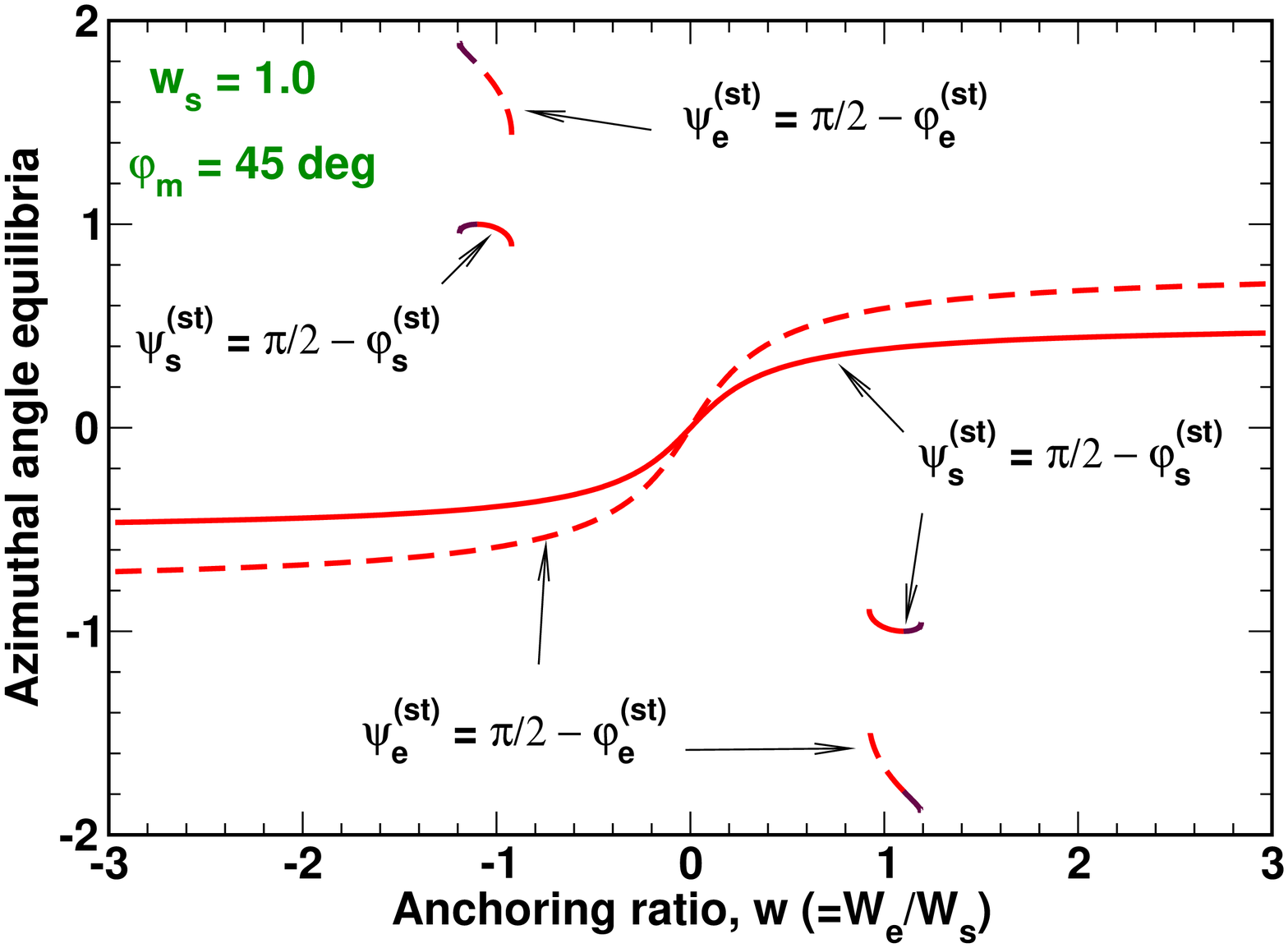}}
\label{subfig:phim-45}
}
\caption{%
Bifurcation diagram for equilibria
of easy axis and surface director azimuthal angles
as a function of the anchoring ratio, 
$w=W_e/W_s$, at  $w_s=1.0$.
Four cases are shown:
(a)~$\varphi_m=\pi/18$;
(b)~$\varphi_m=\pi/9$;
(c)~$\varphi_m=5\pi/36$;
and (d)~$\varphi_m=\pi/4$.
}
\label{fig:bifur-phim}
\end{figure}

%%%%%%%%%%%%
\subsubsection{Bifurcations of equilibria}
\label{subsubsec:bifur-torque}
%%%%%%%%%%%

The surface director
equilibria are represented by
the solutions 
of Eq.~\eqref{eq:D-mu-nu}
that meet the stability conditions~\eqref{eq:stab-cond}.
These conditions combined with the formula~\eqref{eq:stab-matr-Lmb}  
for the matrix $\bs{\Lambda}$ which 
enters the linearization matrix~\eqref{eq:stab-matr-H}
can be analyzed using elementary methods
to yield the following results:
(a)~$D_{++}$ is the branch of stable stationary points (equilibria);
(b)~the branch $D_{--}$ is unstable;
and (c)~for the two remaining branches $D_{\nu\mu}$
with $\nu \mu=-1$,   $D_{-+}$ and $D_{+-}$,
the stability conditions~\eqref{eq:stab-cond} 
are satisfied if and only if
the determinant of the matrix $\bs{\Lambda}_{\nu\mu}$
is positive,
$\det\bs{\Lambda}_{\nu\mu}>0$.
Thus the inequality
\begin{align}
&
\label{eq:stab-cond-mu-nu}
    \nu\sqrt{\bigr[w_s/W_{\mu}^{(1)}(t;w_s)\bigl]^2-t^2} 
\Bigl(
1+2\mu\sqrt{w_s^2-t^2}
\Bigr)
+ \mu\sqrt{w_s^2-t^2}>0
  \end{align}
is the only stability condition 
that defines equilibria belonging to the branches
$D_{-+}$ and $D_{+-}$.

The bifurcation curves
shown in Figs.~\ref{fig:bifur-ws} and~\ref{fig:bifur-phim}
represent the surface director and easy axis 
equilibrium angles computed using 
the parametrizations given in Eqs.~\eqref{eq:bifur-curv-psi-s}
and~\eqref{eq:bifur-curv-psi-e}, respectively.
The numerical procedure involves two steps:
(a)~solving equation~\eqref{eq:D-mu-nu}
at $(\nu,\mu)\in\{(+,+),(+,-),(-,+)\}$;
and (b)~isolating the stable points of $D_{\nu\mu}$ with $\nu\mu=-1$ 
from the unstable ones
based on the stability criterion~\eqref{eq:stab-matr-Lmb}.

Figure~\ref{fig:bifur-ws} demonstrate
the effect of the anchoring parameter
$w_s$ on the bifurcation curves 
that describe dynamical behavior of the model~\eqref{eq:torq-system} 
in relation the anchoring ratio $w=W_e/W_s$
when the phase shift is zero, $\varphi_m=0$.  
The case of small anchoring parameter 
with $w_s<1/2$ was studied in our previous
paper~\cite{Dubtsov:pre:2010}
and is illustrated in Figs.~\ref{subfig:ws-01}--\ref{subfig:ws-049}.

It is seen that there is the only stationary value
of the surface director angle, $\varphi_s^{(\ind{st})}=\pi/2$
($\psi_s^{(\ind{st})}=0$),
provided that $w<w_{+}$ or $w>w_{-}$, where 
$w_{\pm}=1/(1\pm 2 w_s)$ is the critical (bifurcation) value of the anchoring ratio.
The equilibrium values of the easy axis angle in these two regions are:
$\varphi_e^{(\ind{st})}=\pi/2$ 
($\psi_e^{(\ind{st})}=0$ )
at $w<w_{+}$ and $\varphi_e^{(\ind{st})}=0$
($\psi_e^{(\ind{st})}=\pi/2$) 
at $w>w_{-}$.
The latter implies that, 
at large coupling parameter $W_e$ when
$w>w_{-}$, 
the easy axis gliding is completely suppressed,
whereas, in the regime of weak coupling with $w<w_{+}$,
the easy axis rotates  approaching   the stationary state of the
surface director.

Referring to Figs.~\ref{subfig:ws-01}--\ref{subfig:ws-049},
when the anchoring ratio $w$ passes through the critical points
$w_{\pm}$ 
the stationary state $\varphi_{s}^{(\ind{st})}=\pi/2$ becomes unstable
and the pitchfork bifurcations~\cite{Gucken:bk:1990} occur.
So, for the surface director angle in the intermediate region with
 $w_{+}<w<w_{-}$, 
there are two symmetrically arranged stable 
stationary points.
In this case, gliding is not suppressed, but,
by contrast to the regime of weak coupling,
the equilibrium states of the easy axis and the surface director 
are no longer identical.  

Now we examine the important case 
of large anchoring parameter $w_s$
with $w_s>1/2$ 
that represents 
the regime of weak electric field (low voltage).
At $w_s=1/2$,
the largest critical value of the anchoring ratio
$w_{-}$ diverges and 
it can be expected that
bifurcation diagrams at $w_s>1/2$
are characterized by
the only point of pitchfork bifurcation
located at $w=w_{+}$.
An example of such diagram is depicted in Fig.~\ref{subfig:ws-10}.

In the zero-field limit with $E=0$,
the electric coherence length $\xi$
and the parameter $w_s$
both become infinitely large,
whereas the bifurcation point $w_{+}$ decays to zero.
From the formula~\eqref{eq:torq-phie-sol},
it is clear that the equilibrium angles
at $w<w_{+}=0$ ($W_e<0$)
and $w>w_{+}=0$ ($W_e>0$)
are
$\varphi_e^{(\ind{st})}=\varphi_s^{(\ind{st})}=\pi/2$
($\psi_e^{(\ind{st})}=\psi_s^{(\ind{st})}=0$)
and
 $\varphi_e^{(\ind{st})}=\varphi_s^{(\ind{st})}=0$
($\psi_e^{(\ind{st})}=\psi_s^{(\ind{st})}=\pi/2$),
respectively.
So, reorientation of both the surface director and the easy
axis is inhibited provided that 
the coupling constant $W_e$ is positive.

Application of the electric field facilitates the process of reorientation.
At non-vanishing voltage,  
the threshold value of the anchoring ratio
is positive,
$w_{+}>0$, and increases
with electric field.
As is shown in Fig.~\ref{subfig:ws-10},
in the region of strong easy axis coupling
where $w>w_{+}$,
the angles
$\psi_s^{(\ind{st})}$ and $\psi_e^{(\ind{st})}$
differ in magnitude, $|\psi_s^{(\ind{st})}| < |\psi_e^{(\ind{st})}|$.
So, even at large values of the anchoring  ratio
$w$, reorientation of the surface director takes place
as its magnitude is below $\pi/2$.
The latter is not the case for the easy axis
because  the magnitude of the easy axis angle
$\psi_e^{(\ind{st})}$ turned out to be very close to $\pi/2$. 

From the symmetry relation~\eqref{eq:torq-symm-b}
it follows that, at $\varphi_m=0$,
bifurcation diagrams are invariant
under reflection with respect to the $w$-axis: 
$(\psi_s^{(\ind{st})},\psi_e^{(\ind{st})})\to(-\psi_s^{(\ind{st})},-\psi_e^{(\ind{st})})$
(see Fig.~\ref{fig:bifur-ws}).
Generally, this is no longer the case
for nonzero phase shift.

Another important consequence of the symmetry relations~\eqref{eq:torq-symm}
is that under the action of inversion in the origin of coordinates
$(w,\psi_s^{(\ind{st})},\psi_e^{(\ind{st})})\to(-w,-\psi_s^{(\ind{st})},-\psi_e^{(\ind{st})})$
the diagram computed at
$\varphi_m=\pi/4-\Delta\varphi_m$
transforms into the one with $\varphi_m=\pi/4+\Delta\varphi_m$.
Hence we can restrict ourselves to
a set of the diagrams with the phase shift
ranged between $0$ and $\pi/4$, $\varphi_m\in [0,\pi/4]$.

The bifurcation curves presented in Fig.~\ref{fig:bifur-phim}
illustrate the effect of the phase shift on the bifurcation
diagrams at $w_s=1$.
As is evident from Fig.~\ref{subfig:phim-10},
small variations of the phase shift
play the role of a perturbation
that splits the curves at  
the bifurcation point
leading to the formation of
the pattern consisting of two separated
branches of equilibria. 
Such pattern is characteristic of
perturbed (imperfect) pitchfork bifurcations~\cite{Iooss:bk:1990,Crawford:rmp:1991}.

From Figs.~\ref{subfig:phim-20}--\ref{subfig:phim-45},
it can be seen that,
when the phase shift increases,
the low-lying branch of equilibria
shrinks and, in the upper half of the diagram, 
additional branch develops. 
This branch grows with the phase shift
up to the endpoint
$\varphi_m=\pi/4$ of the interval $[0,\pi/4]$.
As is illustrated in Fig.~\ref{subfig:phim-45},
at  this point,
the bifurcation diagram is center symmetric
and 
the bifurcation curves remain intact under
the action of inversion:
$(w,\psi_s^{(\ind{st})},\psi_e^{(\ind{st})})\to(-w,-\psi_s^{(\ind{st})},-\psi_e^{(\ind{st})})$.

\begin{figure}[!tbh]
\centering
\resizebox{120mm}{!}{\includegraphics*{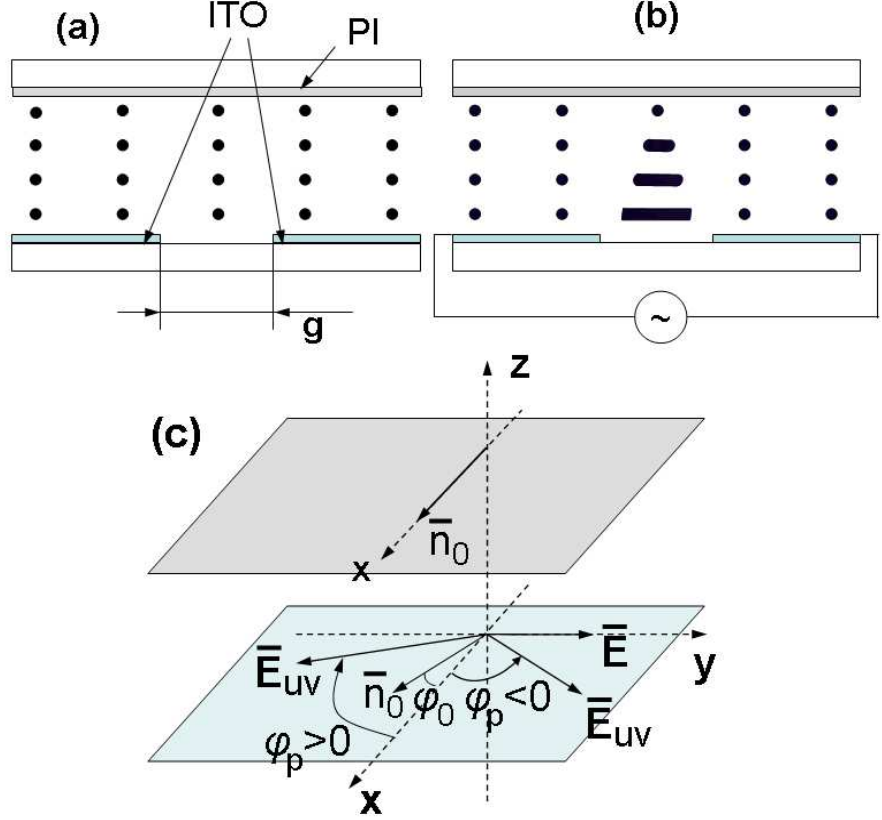}}
\caption{%
Geometry of the experiment: 
(a)~initial state with $\vc{n}_0=\uvc{x}$;
(b)~switching on an in-plane electric field, $\vc{E}=E\uvc{x}$,  
combined with LPUV irradiation; 
(c) the system of reference:
the polarization vector $\vc{E}_{UV}$
of the LPUV light wave propagating 
along the normal to the substrate
(the $z$ axis) is characterized by 
the polarization azimuth $\varphi_p$.
}
\label{fig:expt_geom}
\end{figure}

%%%%%%%%%%%%%%%%%%%%%%%
\section{Experiment}
\label{sec:experiment}
%%%%%%%%%%%%%%%%%%%%%%

In our experiments, liquid crystal (LC) cells ($d=17.4 \pm 0.2\,\mu$m) 
of sandwich like type were assembled 
between two amorphous glass plates. 
The upper glass plate was
covered with a rubbed polyimide film 
to yield the strong planar anchoring conditions.
In Fig.~\ref{fig:expt_geom},
the direction of rubbing
gives the easy axis
parallel to the $x$ axis.
 
A film of the azobenzene sulfonic dye SD1 (Dainippon Ink and
Chemicals)~\cite{Chigrin:bk:2008} 
was deposited onto the bottom substrate
on which transparent indium tin oxide (ITO) electrodes were placed.
The electrodes and the interelectrode stripes
(the gap was about  $g=50~\mu$m) were arranged to be
parallel to the $x$ axis
[see Fig.~\ref{fig:expt_geom}(a)]. 

As in~\cite{Pasechnik:lc:2008,Dubtsov:pre:2010},
the azo-dye SD1 layer was initially illuminated by linearly polarized UV
light (LPUV) at the wavelength $\lambda = 365$~nm.
The preliminary irradiation produced the zones of different energy dose exposure 
$D_p=0.27,\,0.55$~J/cm$^2$ characterized by relatively 
weak azimuthal anchoring strength. 
The light propagating along 
the normal to the substrates (the $z$ axis) 
was selected by an interference filter.
Orientation of the polarization vector of UV light,
$\mathbf{E}_{0}$, was chosen so as to
align azo-dye molecules at a small angle of 4 degrees to 
the $x$ axis, $\varphi_0\approx 4$~deg 
[see Fig.~\ref{fig:expt_geom}(c)].

The LC cell was filled with the nematic LC mixture E7 (Merck) 
in isotropic phase and then slowly cooled down to 
room temperature. 
Thus we prepared the LC cell
with a weakly twisted planar orientational structure 
where the director at the bottom surface 
$\mathbf{n}_0$ is clockwise rotated  
through the initial twist angle $\varphi_0\approx 4$~deg
which is the angle between $\mathbf{n}_0$ 
and the director at the upper substrate (the $x$ axis).

As is indicated in figure~\ref{fig:expt_geom}(b),
the director field deforms when
the in-plane ac voltage ($U=100$~V, $f=3$~kHz)
is applied to the electrodes. 
In addition to the electric field,
$E=2$~V/$\mu$m,
the cell was irradiated with the reorienting
LPUV light beam 
($I_{UV}=0.26$~mW/cm$^2$ and $\lambda = 365$~nm)
normally impinging onto the bottom substrate.

For this secondary LPUV irradiation,
orientation of the polarization plane is determined by
the \textit{polarization azimuth}, 
$\varphi_p$, which is defined as the angle between 
the polarization vector of UV light, $\mathbf{E}_{UV}$, 
and the $x$ axis.
As is indicated in Fig.~\ref{fig:expt_geom}(c),
we shall assume that
positive (negative) values of the polarization azimuth,
$\varphi_p>0$ ($\varphi_p<0$),
correspond to clockwise (counterclockwise) rotation 
of the polarizer from the $x$ axis to $\mathbf{E}_{UV}$.

\begin{figure}%[!tbh]
\centering
   \resizebox{100mm}{!}{\includegraphics*{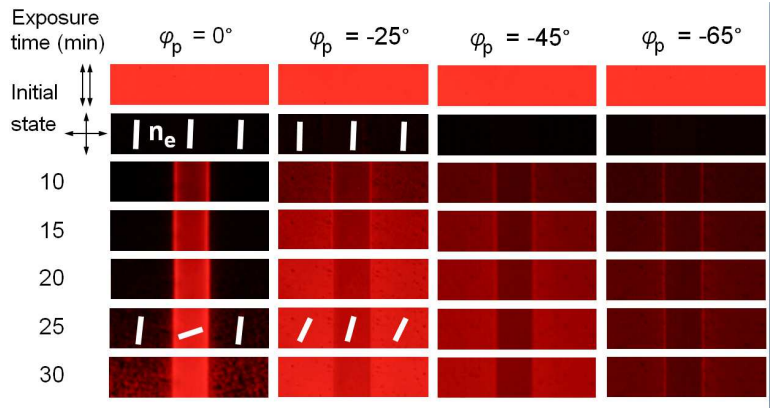}}
\caption{%
Microscopic images of the cell (filter with $\lambda =630$~nm was used)
in crossed polarizers for LPUV irradiation at different
exposure times for various values of the polarization azimuth. 
The interelectrode gap ($g=50\,\mu$m) is indicated.
The ac electric field is $E=2$~V/$\mu$m\,
and the initial irradiation dose is $D_p=0.27$~J/cm$^2$.
}
\label{fig:expt_img}
\end{figure}

Our experimental method has already been described 
in~\cite{Pasechnik:lc:2008,Dubtsov:pre:2010,Kiselev:apl:1:2012}.
In this method, NLC orientational structures 
were observed via a polarized microscope connected
with a digital camera and a fiber optics spectrometer. 
The rotating polarizer technique was used to measure 
the azimuthal angle
$\varphi_e$ characterizing orientation of the easy axis. 
In order to register microscopic images and 
to measure the value of $\varphi_e$,
the electric field and the reorienting light were switched off 
for about 1 min.
This time interval is short enough 
to ensure that orientation of the easy axis remains essentially intact
in the course of measurements.
The measurements were carried out at a temperature of 26$^{\circ}$C.

When the electric field ($E=2$~V/$\mu$m) 
in combination with reorienting LPUV light of the intensity
$I_{UV}=0.26$~mW/cm$^2$ is applied for more than 120 minutes,
we observed the memory effect. 
In this case, after switching off the field and light,
the easy axis did not relax back to its initial state for at least few months.

%%%%%%%%%%%%%%%%%%%%%%%
\subsection{Results}
\label{subsec:results}
%%%%%%%%%%%%%%%%%%%%%%

Figure~\ref{fig:expt_img} shows the microscopic images
obtained at various times of irradiation  
by reorienting LPUV light for four different values of the
polarization azimuth:
$\varphi_p=0$, $-24$, $-45$, $-65$~degrees.
In this case, the initial irradiation dose
is fixed at $D_p=0.27$~J/cm$^2$.  

It can be seen that, 
when the reorienting light is linearly polarized
along the initial surface director $\mathbf{n}_0$ and $\varphi_p=0$,
the brightness of stripes within 
the interelectrode gaps is much higher
as compared to the ones in the region outside the gaps 
where the electric field is negligibly small $E\approx 0$~V/$\mu$m.
So, in  the case of
vanishing polarization azimuth
studied in Refs.~\cite{Pasechnik:lc:2008,Dubtsov:pre:2010},
we arrive at the conclusion that, by contrast
to the electrically assisted light-induced gliding, 
the purely photoinduced reorientation
is almost entirely inhibited.

\begin{figure}
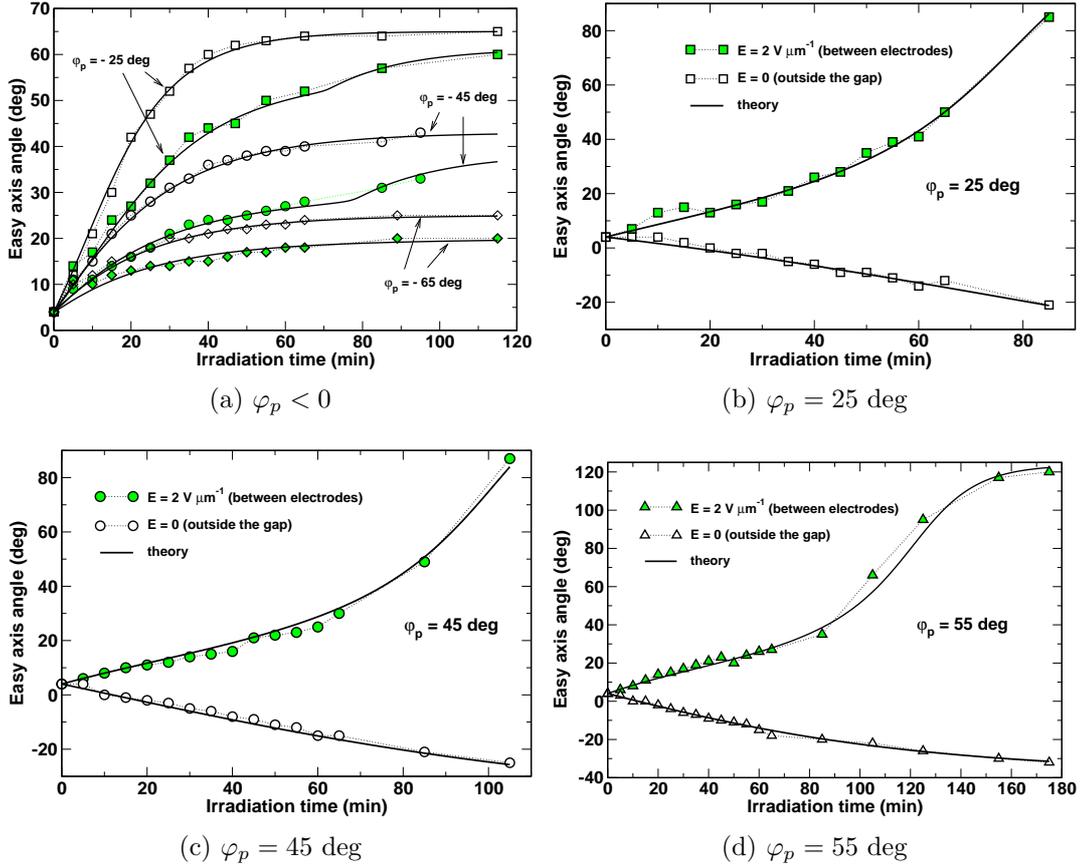
%[!tbh]
\centering
\subfloat[$\varphi_p<0$]{
   \resizebox{69mm}{!}{\includegraphics*{fig5a.eps}}
\label{subfig:phip_neg}
}
\subfloat[$\varphi_p=25$~deg]{
   \resizebox{69mm}{!}{\includegraphics*{fig5b.eps}}
\label{subfig:phip_25}
}
\\
\subfloat[$\varphi_p=45$~deg]{
   \resizebox{69mm}{!}{\includegraphics*{fig5c.eps}}
\label{subfig:phip_45}
}
\subfloat[$\varphi_p=55$~deg]{
   \resizebox{69mm}{!}{\includegraphics*{fig5d.eps}}
\label{subfig:phip_55}
}
\caption{%
Easy axis angle as a function of 
the irradiation time measured
for the reorienting LPUV light
with different 
values of the polarization azimuth $\varphi_p$:
(a)~$\varphi_p<0$ ($D_p=0.27$~J/cm$^2$)
and (b)-(d)~$\varphi_p>0$ ($D_p=0.55$~J/cm$^2$). 
Open (solid) circles, squares and diamonds
represent the data measured in the regions within (outside)
the interelectrode gaps
where $E=2$~V/$\mu$m\, ($E\approx 0$~V/$\mu$m).
The results computed from the phenomenological model by solving equations~\eqref{eq:phi_e-orig}
and~\eqref{eq:phi_s-orig}
are shown as solid lines.
}
\label{fig:expt_data}
\end{figure}

The latter is no longer the case
for the reorienting light with nonzero polarization azimuth.
Referring to Fig.~\ref{fig:expt_img},
at $\varphi_p\ne 0$, 
light-induced distortions of the surface director in 
the zero-field region located outside the gaps 
are very much more pronounced.
It is also evident from the curves
depicted in Fig.~\ref{fig:expt_data}(a)
representing the irradiation time dependencies of 
the easy axis angle measured
at negative polarization azimuthal angles of 
the reorienting LPUV light. 

In the zero-field curves,
the easy axis angle increases with
the irradiation time
starting from the angle of initial twist,
$\varphi_0$,
and
approaches the photo-steady state
characterized by the photosaturated value of the
angle close to $\pi/2+\varphi_p$. 
The curves describing the electrically assisted
reorientation within the interelectrode gaps
lie below the zero-field ones and
reveal analogous behavior.

The data measured at 
the initial irradiation dose $D_p=0.55$~J/cm$^2$.
for the
polarization azimuths of the opposite sign,
$\varphi_p>0$, (see Fig.~\ref{fig:expt_data}(b)-(d))
show that, in the zero-field region,
the light-induced changes of the easy axis angle 
are negative and correspond to the counterclockwise
rotation of the polarizer.
As seen from Fig.~\ref{fig:expt_data}(b)-(d),
the dynamics of the easy axis in the presence of the electric field
essentially differs from the one in the regime of purely photoinduced
reorientation.

\begin{figure}%[!tbh]
\centering
   \resizebox{90mm}{!}{\includegraphics*{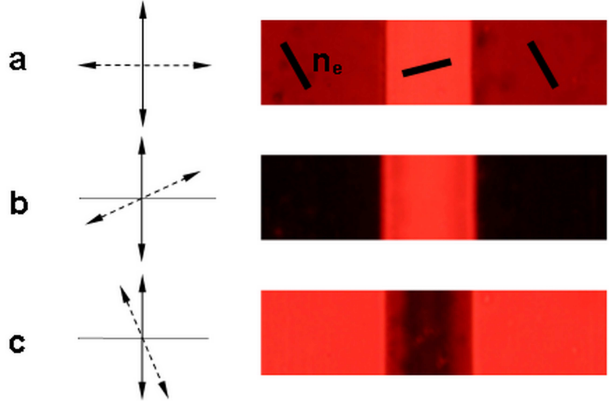}}
\caption{%
Microscopic images
obtained at $\varphi_p>0$
for different orientations of the polarizer:
(a)~crossed polarizers;
dark state orientation for the regions
(b)~outside and (c)~within the interelectrode gaps. 
}
\label{fig:expt_pol}
\end{figure}

In the interelectrode gaps,
it turned out that
the electric field prevails thus suppressing 
the tendency for the easy axis to be reoriented
along the normal to the polarization vector of light 
$\mathbf{E}_{UV}$.
This effect is illustrated in figure~\ref{fig:expt_pol}
that, in particular, 
presents the microscopic images 
registered in the dark states
of both the zero-field and the interelectrode 
regions. 
It is indicated that
the azimuthal angles describing 
orientation of the polarizer (and of the easy axis)
in these states are opposite in sign.

%%%%%%%%%%%%
\subsection{Model versus experiment}
\label{subsec:torq-model-expt}
%%%%%%%%%%%

Now we briefly discuss how
the experimental data can be interpreted 
using the above phenomenological model.
According to this model, the dynamics of the easy axis and surface
director azimuthal angles, $\varphi_e$ and $\varphi_s$,
is governed by the system of balance torque
equations~\eqref{eq:torq-orig}.
 
We can apply the formula~\eqref{eq:torq-phie-sol} to 
fit the experimental data measured outside the interelectrode gaps.
For $\varphi_p<0$ and  $D_p=0.27$~J/cm$^2$, the theoretical curves
presented in figure~\ref{fig:expt_data}(a) are computed at $t_e\approx
23$~min. The values of $\varphi_m$ are:
$\varphi_m=25$~deg at $\varphi_p=-65$~deg;
$\varphi_m=43$~deg at $\varphi_p=-45$~deg;
and $\varphi_m=65$~deg at $\varphi_p=-25$~deg.
Similarly, for the case where $\varphi_p>0$ and  $D_p=0.55$~J/cm$^2$,
the results shown in figure~\ref{fig:expt_data}(b) are computed at $t_e\approx
85$~min. In this case, the values of $\varphi_m$ are:
$\varphi_m=-38$~deg at $\varphi_p=55$~deg;
$\varphi_m=-43$~deg at $\varphi_p=45$~deg;
and $\varphi_m=-65$~deg at $\varphi_p=25$~deg.

In the presence of electric field (the region between the
electrodes), the dynamical system~\eqref{eq:torq-orig}
has to be
solved numerically. 
As in the previous section,
it  is convenient to work with the
system rewritten in the dimensionless form~\eqref{eq:torq-system} 
which is
characterized by the characteristic time
$\tau_e=\gamma_e/K_e$, the viscosity ratio $\gamma=\gamma_e/\gamma_s$ 
and the two dimensionless anchoring parameters: 
$w_e=W_e/(2 K_E)$ and $w_s=W_s/(2 K_E)$.

The parameters used for computing the curves shown in
figure~\ref{fig:expt_data}(a) ($\varphi_p<0$ and
$D_p=0.27$~J/cm$^2$): $\tau_e=230$~min, $\gamma=100$
(the gliding viscosity is typically several orders higher than the
surface viscosity), $w_e=5$ and $w_s=1$.
The values of $\varphi_m$ are:
$\varphi_m=25.1$~deg at $\varphi_p=-65$~deg;
$\varphi_m=35$~deg at $\varphi_p=-45$~deg;
and $\varphi_m=60$~deg at $\varphi_p=-25$~deg.
In figure~\ref{fig:expt_data}(b) representing the case
where $\varphi_p>0$ and
$D_p=0.55$~J/cm$^2$, the theoretical results are calculated
at $\tau_e\approx 39.2$~min, $\gamma=100$, 
$w_e=1.1$, $w_s=10$ and the values of $\varphi_m$
listed as follows:
$\varphi_m=-41$~deg at $\varphi_p=55$~deg;
$\varphi_m=-45$~deg at $\varphi_p=45$~deg;
and $\varphi_m=-50$~deg at $\varphi_p=25$~deg.

As it can be seen from figures~\ref{fig:expt_data}(a)-(b), 
the computed curves are in good agreement
with the experimental data. The results of fitting
indicate that the initial irradiation dose
has a profound effect on the anchoring parameters, 
whereas the electric field
affects the polarization dependent phase shift $\varphi_m$.

%%%%%%%%%%%%%%%%%
\section{Discussion and conclusions}
\label{sec:conclusion}
%%%%%%%%%%%%%%%%

In conclusion, we have experimentally studied 
the effects of polarization azimuth
in the electrically assisted
light-induced azimuthal gliding of the NLC easy axis 
on the photoaligning azo-dye film. 
It is found that,
by contrast to the case where
the polarization vector is oriented
along the initial surface director,
at nonzero polarization angle 
$\varphi_0$,
the purely photoinduced reorientation takes place
outside the interelectrode gaps.
For such field-free regime of reorientation,
the easy axis reorients 
approaching the photosaturation limit
close to the normal to the polarization vector.
These results agree with 
the theoretical predictions
of the diffusion model~\cite{Kis:pre:2009}
describing kinetics of photoinduced ordering in azo-dye films
and the corresponding theoretical analysis will be published 
in a  separate paper.
 
In the regions between electrodes
with non-vanishing electric field,
the dynamics of reorientation slows down with
the polarization azimuth
and, as opposed to the case of 
purely photoinduced reorientation,
the sense of easy axis rotation
for the electrically assisted light-induced gliding 
is found to be independent of the sign of polarization azimuth. 
The above mentioned theory~\cite{Kis:pre:2009}
cannot be directly applied to this case
and we have shown that 
the phenomenological model~\cite{Pasechnik:lc:2006,Pasechnik:lc:2008}
can be extended to interpret our data. 

\begin{acknowledgments} 
This work was partially supported by grants: RF Ministry of Education
and Science: 14.B37.21.0894, 14.B37.21.1198, 14.B37.21.1914;
CERG Grant 612310 and 612409.
\end{acknowledgments}

%\bibliographystyle{apsrev}
%\bibliography{optics,polymer,scatter,lc,quant,hk,flc,qft,math,my_papers}


\end{document}